\documentclass[prl,twocolumn,showkeys,
nofootinbib,
 amsmath,amssymb,
 aps,
]{revtex4-2}
\usepackage{yfonts}
\usepackage{graphicx}
\usepackage{dcolumn}
\usepackage{bm}
\usepackage{nicefrac}
\usepackage{color}
\definecolor{darkblue}{RGB}{0,0,196}
\definecolor{darkgreen}{RGB}{0,120,0}
\usepackage[colorlinks=true,linktocpage=true,linkcolor=darkblue,citecolor=red,urlcolor=darkblue]{hyperref}
\usepackage{hyperref}

\newcommand{\bea}{\begin{eqnarray}}
\newcommand{\eea}{\end{eqnarray}}
\newcommand{\bel}[1]{\begin{eqnarray}\label{#1}}
\newcommand{\eel}{\end{eqnarray}}

\def\LB{\left(}
\def\RB{\right)}
\def\LSB{\left[}
\def\RSB{\right]}

\newcommand{\nn}{\nonumber}

\newcommand{\EQ}[1]{Eq.~(\ref{#1})}
\newcommand{\EQn}[1]{(\ref{#1})}

\newcommand{\EQSTWO}[2]{Eqs.~(\ref{#1})~and~(\ref{#2})}
\newcommand{\EQSTWOn}[2]{(\ref{#1})~and~(\ref{#2})}

\newcommand{\EQSM}[2]{Eqs.~(\ref{#1})--(\ref{#2})}
\newcommand{\EQSMn}[2]{(\ref{#1})--(\ref{#2})}

\newcommand{\CIT}[1]{Ref.~\citep{#1}} 
\newcommand{\CITn}[1]{\citep{#1}} 

\newcommand{\p}{\partial}

\def\epsUabgd{\epsilon^{\alpha \beta \gamma \delta}}

\newcommand{\av}{{\boldsymbol a}} 
\newcommand{\bv}{{\boldsymbol b}} 
 
\newcommand{\kv}{{\boldsymbol k}}

\newcommand{\pv}{{\boldsymbol p}}
\newcommand{\tv}{{\boldsymbol t}}

\newcommand{\sv}{{\boldsymbol s}}

\newcommand\omv{{\boldsymbol \omega}}

\newcommand{\f}[2]{\frac{#1}{#2}}
\newcommand{\onehalf}{{\nicefrac{1}{2}}} 
 
\newcommand{\threefourths}{{\nicefrac{3}{4}}} 

\def\spin{\,\textgoth{s:}}
\def\spinl{|{\boldsymbol s}_*|}

\begin{document}


\title{Generalized thermodynamic relations for perfect spin hydrodynamics}

\author{Wojciech Florkowski}
\email{wojciech.florkowski@uj.edu.pl}
\affiliation{Institute of Theoretical Physics, Jagiellonian University, PL-30-348 Krak\'ow, Poland}

\author{Mykhailo Hontarenko}
\email{mykhailo.hontarenko@student.uj.edu.pl}
\affiliation{Institute of Theoretical Physics, Jagiellonian University, PL-30-348 Krak\'ow, Poland}

\date{\today}

\begin{abstract}
Generalized thermodynamic relations are introduced into the framework of a relativistic perfect spin hydrodynamics. They allow for consistent treatment of spin degrees of freedom, including the use of spin tensors whose structure follows from microscopic calculations. The obtained results are important for establishing consistency between different formulations of spin hydrodynamics and form the basis for introducing dissipative corrections.
\end{abstract}

                              
\maketitle

\noindent {\it Introduction ---} Recent measurements of non-zero spin polarization of hyperons \cite{STAR:2017ckg, STAR:2018gyt, STAR:2019erd} and vector mesons \cite{ALICE:2019aid} produced in relativistic heavy-ion collisions have triggered broad interest in the spin polarization phenomena in strongly interacting matter, for a recent review see Ref.~\cite{Niida:2024ntm}. On the theory side, there exist several approaches to incorporate spin degrees of freedom into the framework of relativistic hydrodynamics ~\CITn{Becattini:2009wh,  Becattini:2011zz, Montenegro:2017rbu, Florkowski:2017ruc, Hattori:2019lfp, Weickgenannt:2019dks, Bhadury:2020puc, Li:2020eon, Weickgenannt:2020aaf, Bhadury:2020cop, Fukushima:2020ucl, Shi:2020htn, Montenegro:2020paq, Weickgenannt:2021cuo,  Gallegos:2021bzp, Hongo:2021ona, Singh:2022ltu, Hu:2021pwh, She:2021lhe, Weickgenannt:2022zxs, Wagner:2022amr, Kumar:2023ojl,Weickgenannt:2023nge, Wagner:2024fhf}. The latter has become the main theoretical tool used to describe the spacetime evolution of strongly interacting matter produced in heavy-ion collisions~\CITn{Florkowski:2010zz,Gale:2013da,Jaiswal:2016hex}, hence, the inclusion of spin dynamics in the hydrodynamics formalism seems to be an inevitable necessity.

A certain difficulty in developing the formalism of spin hydrodynamics is the fact that there are different formulations of this approach using different assumptions. The differences appear already at the basic level of relativistic thermodynamic relations and definitions of the fundamental macroscopic quantities such as the spin tensor. 

In this work we critically reexamine thermodynamic relations used in perfect spin hydrodynamics of particles with spin $\onehalf$ and propose to introduce their generalized (tensor) forms that can be used for large values of the spin polarization tensor, $\omega_{\mu\nu}$, and with kinetic-theory motivated forms of the spin tensor $S^{\lambda, \mu\nu}$. In this way, we remove a gap between the works that use kinetic-theory concepts as the starting point~\CITn{Florkowski:2017dyn,Florkowski:2018ahw} and the works that use phenomenological expressions for the spin tensor and construct dissipative corrections using the positivity of the entropy production as the main physical ansatz~\CITn{Hattori:2019lfp,Fukushima:2020ucl}. 

Interestingly, the new tensor forms of the thermodynamic relations proposed in this work include terms whose mathematical structure is typical for dissipative corrections. In our case their presence is not related to the entropy production but results from a richer description of the system that requires introduction of the spin polarization tensor (reminding us of the structure of anisotropic relativistic magnetohydrodynamics (MHD), for example, see~\CITn{PhysRevE.51.4901}). An important consequence of the fact that such terms appear at the perfect-fluid level is that they should be taken into account in theoretical constructions aiming at the development of dissipative spin hydrodynamics.

\smallskip
\noindent
{\it Notation and conventions ---} For the Levi-Civita tensor $\epsilon^{\mu\nu\alpha\beta}$ we follow the convention $\epsilon^{0123} =-\epsilon_{0123} = +1$. The metric tensor is of the form $g_{\mu\nu} = \textrm{diag}(+1,-1,-1,-1)$. Throughout the text we make use of natural units, $\hbar = c = k_B = 1$. The scalar product of two four-vectors $a$ and $b$ reads $a \cdot b = a^0 b^0 - \av \cdot \bv$, where the three-vectors are denoted by bold font.

\smallskip
\noindent {\it Scalar and tensor forms of thermodynamic relations ---} The fundamental thermodynamic relations used in standard relativistic hydrodynamics consist of the identity
\bel{eq:ext_S}
\varepsilon + P  = T \sigma + \mu n
\eel 
and the first law of thermodynamics
\bel{eq:firstlaw_S}
d\varepsilon =Td\sigma + \mu dn .
\eel 
Here $\varepsilon$, $P$, $T$, $\sigma$, $\mu$ and $n$ are the local energy density, pressure, temperature, entropy density, baryon chemical potential, and baryon number density, respectively. The identity \EQn{eq:ext_S} is a direct consequence of the extensivity of energy, entropy and baryon number (they are all proportional to the system's volume). Equations \EQSTWOn{eq:ext_S}{eq:firstlaw_S} imply the Gibbs-Duhem relation
\bel{eq:GD_S}
dP = \sigma dT + n d\mu .  
\eel 
In order to take into account dissipation effects, one usually rewrites \EQSM{eq:ext_S}{eq:firstlaw_S} in a tensor (four-vector) form. This is achieved by multiplication of \EQSM{eq:ext_S}{eq:GD_S} by the local four-velocity of the fluid $u^\mu$, which leads to the following expressions
\bel{eq:ext_T}
S^\mu = \sigma u^\mu = P \beta^\mu - \xi N^\mu + \beta_\lambda T^{\lambda\mu},
\eel
\bel{eq:firstlaw_T}
dS^\mu = - \xi dN^\mu + \beta_\lambda dT^{\lambda\mu},
\eel
\bel{eq:GD_T}
d(P\beta^\mu) = N^\mu d\xi - T^{\lambda\mu} d\beta_\lambda.
\eel
Here we have introduced common notation: $\beta^\mu = u^\mu/T$, $\beta = \sqrt{\beta^\lambda \beta_\lambda} = 1/T$, and $\xi = \mu/T$. The tensors $N^\mu$ and $T^{\lambda\mu}$ describe the baryon current and energy-momentum tensor for a perfect fluid, namely, $N^\mu = n u^\mu$ and $T^{\lambda\mu} = (\varepsilon+P) u^\lambda u^\mu - P g^{\lambda \mu} = \varepsilon u^\lambda u^\mu - P \Delta^{\lambda \mu}$, where the tensor $\Delta^{\lambda \mu} = g^{\lambda \mu}-u^\lambda u^\mu$ projects on the space orthogonal to flow.

Many formulations of relativistic spin hydrodynamics as their starting points choose an extension of \EQSM{eq:ext_S}{eq:GD_S} that includes the spin polarization tensor $\omega_{\alpha\beta}$, the tensor spin chemical potential $\Omega_{\alpha\beta} = T \,\omega_{\alpha\beta}$, and the spin density tensor $S^{\alpha \beta}$. They read
\bel{eq:ext_Ss}
\varepsilon + P  = T \sigma + \mu n + 
{\scriptstyle{ \frac{1}{2} }}
\Omega_{\alpha\beta} S^{\alpha \beta},
\eel 
\bel{eq:firstlaw_Ss}
d\varepsilon =T d\sigma + \mu dn + 
{\scriptstyle{ \frac{1}{2} }}
\Omega_{\alpha\beta} dS^{\alpha \beta},
\eel 
\bel{eq:GD_Ss}
dP = \sigma dT + n d\mu  + 
{\scriptstyle{ \frac{1}{2} }}
S^{\alpha \beta} d\Omega_{\alpha\beta}.  
\eel 
We note that $\omega_{\alpha\beta}$,  $\Omega_{\alpha\beta}$, and $S_{\alpha \beta}$ are all rank-2 antisymmetric tensors. Below we will use the following parametrization of the spin polarization tensor~\CITn{Florkowski:2017ruc}
\bel{eq:ko}
\omega_{\alpha\beta} = k_\alpha u_\beta - k_\beta u_\alpha + t_{\alpha\beta},
\eel 
where $t_{\alpha\beta}=\epsilon_{\alpha\beta\gamma\delta} u^\gamma \omega^\delta$, and the four-vectors $k$ and $\omega$ are orthogonal to the flow vector $u$, namely, $k\cdot u = 0$ and $\omega \cdot u = 0$. By multiplying \EQSM{eq:ext_Ss}{eq:GD_Ss} by $u^\mu$, we obtain
\bel{eq:ext_Ts}
S^\mu =  P \beta^\mu - \xi N^\mu + \beta_\lambda T^{\lambda\mu} - 
{\scriptstyle{ \frac{1}{2} }}
\omega_{\alpha\beta} S^{\mu, \alpha \beta},
\eel
\bel{eq:firstlaw_Ts}
dS^\mu = - \xi dN^\mu + \beta_\lambda dT^{\lambda\mu}
- {\scriptstyle{ \frac{1}{2} }}
\omega_{\alpha\beta} dS^{\mu, \alpha \beta},
\eel
\bel{eq:GD_Ts}
d(P\beta^\mu) = N^\mu d\xi - T^{\lambda\mu} d\beta_\lambda +{\scriptstyle{ \frac{1}{2} }}
 S^{\mu, \alpha \beta} d\omega_{\alpha\beta}.
\eel
Here, we have introduced the spin tensor $S^{\mu, \alpha \beta}$ defined by the expression
\bel{eq:spint_F}
S^{\mu, \alpha \beta} = u^\mu S^{\alpha \beta},
\eel
which is an analog of the perfect-fluid forms of $N^\mu$ and $T^{\lambda\mu}$ given below~\EQ{eq:GD_T}. A direct consequence of \EQ{eq:firstlaw_Ts} is that it implies the entropy conservation for a system that conserves baryon number, energy, linear momentum and spin, namely, the conservation laws $\p_\mu N^\mu =0$, $\p_\mu T^{\mu \lambda}=0$,  and $\p_\mu S^{\mu, \alpha\beta}~=~0$ imply $\p_\mu S^\mu =0$. We note that the spin conservation is a direct consequence of using a symmetric energy-momentum tensor in the considered formalism.

The use of the expression \EQn{eq:spint_F} can be traced back to the very first model of a spinning fluid by Weyssenhoff and Raabe~\CITn{Weyssenhoff:1947iua}. It was also used in~\CIT{Florkowski:2017ruc}, where the first formulation of relativistic hydrodynamics for particles with spin $\onehalf$ was proposed. The form \EQn{eq:spint_F} has been subsequently used in many works that followed the methods of Israel and Stewart (positivity of the entropy current) to construct the framework of dissipative spin hydrodynamics~\CITn{Hattori:2019lfp, Fukushima:2020ucl, Sarwar:2022yzs, Daher:2022wzf, Xie:2023gbo, Daher:2024bah}. 

Although \EQ{eq:spint_F} has been used in numerous works, its form disagrees with expressions for the spin tensor obtained from the microscopic calculations~\CITn{DeGroot:1980dk} (and used in the spin hydrodynamics formulations that directly refer to kinetic theory~\CITn{Florkowski:2017dyn,Florkowski:2018fap}). The latter usually lead to a more complex structures. As an extension of  \EQ{eq:spint_F} we may consider the form
\bel{eq:spint_GLW}
S^{\lambda, \mu \nu} &=& u^\lambda 
\LSB A \LB k^\mu u^\nu - k^\nu u^\mu \RB + A_1 t^{\mu\nu}
\RSB  \\
&& + \frac{A_3}{2} 
\LB t^{\lambda \mu} u^\nu - t^{\lambda \nu} u^\mu + 
\Delta^{\lambda \mu} k^\nu - \Delta^{\lambda \nu} k^\mu 
\RB, \nonumber
\eel
where $A, A_1$ and $A_3$ are some scalar functions. In the case $A_3=0$, we reproduce \EQ{eq:spint_F}. Moreover, for $A=A_1$ (with $A_3=0$) we find that the spin density tensor is proportional to the spin polarization tensor, namely $S^{\mu\nu}= A \, \omega^{\mu\nu}$~\footnote{Recently, the spin equation of state of the form $S^{\mu\nu}= A \, \omega^{\mu\nu}$ has been analysed and excluded \CITn{Daher:2024ixz} as leading to unstable behavior of rest frame modes in the first-order \CITn{Sarwar:2022yzs,Daher:2022wzf} and second-order \CITn{Xie:2023gbo,Daher:2024bah} dissipative spin hydrodynamics.}.

The above discussion indicates an important problem encountered in the formulations of spin hydrodynamics --- a transition from \EQSM{eq:ext_Ss}{eq:GD_Ss} to \EQSM{eq:ext_Ts}{eq:GD_Ts} obtained by the multiplication of  \EQSM{eq:ext_Ss}{eq:GD_Ss} by the flow vector $u^\mu$ is inconsistent with the use of a microscopically derived spin tensor as the latter contains parts orthogonal to $u^\mu$. Consequently, the formulations of spin hydrodynamics that start from \EQSM{eq:ext_Ss}{eq:GD_Ss} and use the spin tensor of the form \EQn{eq:spint_F} seem to be inconsistent with the formulations based on the kinetic-theory arguments.  In this work we argue that the solution to the above problem lies in revising the thermodynamic relations \EQSMn{eq:ext_Ss}{eq:GD_Ss}. Our reasoning is supported by an analysis of a kinetic model presented below.

\medskip
\noindent {\it Insights from kinetic theory ---} Let us turn now to the discussion of a simple kinetic model that treats spin classically. It has been shown that for small polarization tensor the results obtained with such a model are consistent with the results obtained from the calculations using a semiclassical expansion of the Wigner function~\CITn{Florkowski:2018fap}.

In the classical treatment of spin~\CITn{Mathisson:1937zz,2010GReGr..42.1011M}, one introduces the internal angular momentum tensor $s^{\alpha\beta}$ defined in terms of the particle's four-momentum~$p$ (with $p^\mu p_\mu = m^2$ being the particle mass squared) and spin four-vector~$s$
\bel{eq:sab}
s^{\alpha\beta} = \f{1}{m} \epsUabgd p_\gamma s_\delta.
\eel
Equation \EQn{eq:sab} implies that $s^{\alpha\beta} = -s^{\beta\alpha}$ and $s^{\alpha\beta} p_\beta = 0$. The spin four-vector is orthogonal to four-momentum $s \cdot p = 0$, hence we can write $s^{\alpha} = 1/(2m) \,\epsUabgd p_\beta s_{\gamma \delta}$. In the particle's rest frame (PRF), where $p^\mu = (m,0,0,0)$, the four-vector $s^\alpha$ has only space components, $s^\alpha = (0,\sv_*)$, with the normalization $\spinl = \spin$. For particles with spin $\onehalf$ we use the value of the Casimir operator $\spin^2 = \onehalf \left( 1+ \onehalf  \right) = \threefourths$.

The basic object used in the kinetic theory is the phase-space distribution function $f(x,\pv)$. For particles with spin, $f(x,\pv)$ is generalized to a spin dependent distribution $f(x,\pv,s)$. In local equilibrium, the spin dependent distribution functions for particles ($+$) and antiparticles ($-$) have the form~\footnote{In the main text, we restrict our considerations to the classical Boltzmann statistics. The case of the Fermi-Dirac statistics is worked out in the Supplemental Material. It also leads to Eqs.~\EQn{eq:Hmu2}, \EQn{eq:firstlaw_TsG}, and \EQn{eq:GD_TsG}, which indicates universality of our findings. }
\bel{eq:fpm-spin}
f^\pm(x,p,s) = \exp\LB - p_\mu \beta^\mu \pm \xi + 
{\scriptstyle{ \frac{1}{2} }} 
\omega_{\alpha \beta} s^{\alpha\beta} \RB.
\eel 
where $\beta^\mu, \xi$ and $\omega_{\alpha \beta}$ are functions of space and time coordinates $x$ and play the same role as $\beta^\mu, \xi$ and $\omega_{\alpha \beta}$ defined above. By integrating the equilibrium distribution functions over momentum and spin degrees of freedom, one obtains the macroscopic currents and tensors
\bel{eq:Nmu}
N^\mu\!=\!\int dP \,dS \, p^\mu \, \left[f^+(x,p,s)-f^-(x,p,s) \right],
\eel
\bel{eq:Tmunu}
T^{\mu \nu}\!=\!\int dP \,dS \, p^\mu p^\nu \, \left[f^+(x,p,s) + f^-(x,p,s) \right],
\eel 
\bel{eq:Slmunu}
\hspace{-0.5cm}S^{\lambda, \mu\nu}\!&=&\!\!\int \!dP \, dS \, \, p^\lambda \, s^{\mu \nu} 
\left[f^+(x,p,s)+ f^-(x,p,s) \right]. 
\eel
Here we have introduced the integration measures in momentum, $dP = d^3p/((2\pi)^3 E_p)$, and spin space~\CITn{Florkowski:2018fap}
\bel{eq:dS}
dS = \f{m}{\pi \spin}  \, d^4s \, \delta(s \cdot s + \spin^2) \, \delta(p \cdot s),
\eel
with the normalization $\int dS = 2$ that reflects two possible orientations of the spin $\onehalf$.

In addition to $N^\mu$, $T^{\mu \nu}$, and $S^{\lambda, \mu\nu}$, we introduce the entropy current using the standard Boltzmann definition~\CITn{Landau:1980mil}
\bel{eq:Hmu1}
\hspace{-0.75cm} S^\mu =\!-\!\int dP \, dS \, p^\mu 
\LSB 
f^+ \LB \ln f^+\!\!-\!1\RB\!+\! 
f^- \LB \ln f^-\!\!-1\!\RB \RSB,
\eel
which directly leads to the formula~\CITn{Florkowski:2019qdp}
\bel{eq:Hmu2}
S^\mu =  T^{\mu \alpha} \beta_\alpha-\f{1}{2} \omega_{\alpha\beta} S^{\mu, \alpha \beta}
-\xi N^\mu + {\cal N}^\mu,
\eel
where we have defined the particle four-current
\bel{eq:calN}
{\cal N}^\mu = \coth\xi\,\,N^\mu .
\eel

\medskip
\noindent {\it Generalized thermodynamics ---} We reach now the key moment of our discussion. It is important to realize that \EQ{eq:Hmu2} has exactly the same structure as \EQ{eq:ext_Ts} except that the term $P \beta^\mu$ in \EQ{eq:ext_Ts} is replaced by ${\cal N}^\mu$. Obviously, these two tensors agree in the spinless case. Moreover, they also agree if only linear corrections in the spin polarization tensor are included, as the spin effects in both ${N}^\mu$ and ${\cal N}^\mu$ start with the quadratic terms in $k$ and $\omega$. However, in general we have $P \beta^\mu \neq {\cal N}^\mu$. Thus, \EQ{eq:Hmu2} represents a generalization of the standard thermodynamic relation \EQn{eq:ext_Ts} to the case including the spin degrees of freedom. 

Starting from the definition \EQn{eq:calN}, one can derive two additional relations
\bel{eq:firstlaw_TsG}
dS^\mu = - \xi dN^\mu + \beta_\lambda dT^{\lambda\mu}
- {\scriptstyle{ \frac{1}{2} }}
\omega_{\alpha\beta} dS^{\mu, \alpha \beta},
\eel
\bel{eq:GD_TsG}
d{\cal N}^\mu = N^\mu d\xi - T^{\lambda\mu} d\beta_\lambda + {\scriptstyle{ \frac{1}{2} }}
 S^{\mu, \alpha \beta} d\omega_{\alpha\beta}. 
\eel
Equation \EQn{eq:firstlaw_TsG} has the same form as \EQ{eq:firstlaw_Ts}, however, with the spin tensor \EQn{eq:spint_F} replaced by the formula \EQn{eq:spint_GLW}. Equation \EQn{eq:GD_TsG} agrees with \EQ{eq:GD_Ts} only if we can again set $P \beta^\mu = {\cal N}^\mu$. Equation \EQn{eq:firstlaw_TsG} also shows that the entropy conservation is a direct consequence of three other conservation laws: for baryon number, energy, linear momentum, and spin.

A set of Eqs.~\EQn{eq:Hmu2}, \EQn{eq:firstlaw_TsG}, and \EQn{eq:GD_TsG} represents our first important result. For spin polarized media, it should replace the set of Eqs.~\EQSMn{eq:ext_Ts}{eq:GD_Ts}. Strictly speaking, \EQ{eq:Hmu2} was derived for the first time in \CIT{Florkowski:2019qdp}, however, in the subsequent papers only the terms linear in $\omega_{\mu\nu}$ were included that resulted in neglecting all the products (contractions) of the tensors $\omega_{\mu\nu}$ and $S^{\lambda, \mu\nu}$. One should emphasize that \EQ{eq:Hmu2} holds for any values of the spin polarization tensor. Equation \EQn{eq:firstlaw_TsG} was used earlier in the works that derived the form of the dissipative corrections in spin hydrodynamics. However, in this series of investigations, to maintain the products of $\omega_{\mu\nu}$ and $S^{\lambda, \mu\nu}$ in the formalism, it was assumed that $S^{\lambda, \alpha\beta}$ was of the form \EQn{eq:spint_F} with $S^{\alpha\beta}$ being of the zeroth order in $\omega^{\alpha\beta}$. These assumptions contradict the microscopic results which suggest the form \EQn{eq:spint_GLW} with $A_3 \neq 0$. We conclude this part of our discussion with the statement that Eqs.~\EQn{eq:Hmu2}, \EQn{eq:firstlaw_TsG}, and \EQn{eq:GD_TsG} should be used with at least second order corrections in $\omega$ to include the spin degrees of freedom in a non-trivial and consistent way~\footnote{With only linear terms in $\omega^{\mu \nu}$ included, we also obtain a consistent description, however, with a rather trivial treatment of thermodynamic relations which reduce to a spinless case. Then, the spin dynamics is determined by the hydrodynamic background defined by standard hydrodynamic relations.}. 

\medskip
\noindent {\it Generalized scalar thermodynamic relations ---} The arguments presented above indicate that the generalized tensor thermodynamic relations include currents and tensors that contain parts orthogonal to the flow vector~$u^\mu$. Such terms typically appear in dissipative hydrodynamics, however, in the spin hydrodynamics that may appear at the perfect-fluid (entropy conserving) level.

It becomes clear now, that such orthogonal corrections cannot appear if one starts from \EQSM{eq:ext_Ss}{eq:GD_Ss} and multiplies them by $u^\mu$. Hence, a natural question arises, if there exists an analog of such scalar thermodynamic relations that is valid in the case of arbitrary $\omega^{\mu\nu}$ and for microscopically motivated spin tensor of the form \EQn{eq:spint_GLW}. 

In this case the baryon current $N^\mu$ has the structure~\footnote{Explicit calculations supporting the discussed decompositions are given in the supplemential materials where the results of the calculations including second-order corrections in $\omega^{\mu\nu}$ are given.}
\bel{eq:NmuG}
N^\mu = {\bar n} u^\mu + n_t t^\mu,
\eel
where 
\bel{eq:t}
t^\mu = t^{\mu\nu} k_\nu = \epsilon^{\mu \nu \alpha \beta} k_\nu u_\alpha \omega_\beta. 
\eel 
The four-vector $t$ is orthogonal to the vectors $u, k$ and $\omega$. In the local rest frame (LRF), where $u^\mu=(1,0,0,0)$, one finds that $\tv = \kv \times \omv$. The current ${\cal N}^\mu$ can be obtained from \EQ{eq:calN} that always holds for the Boltzmann statistics. In analogy to \EQ{eq:NmuG} we find 
\bel{eq:NSmuG}
S^\mu = {\bar \sigma} u^\mu + \sigma_t t^\mu.
\eel
The scalar functions ${\bar n}$ and ${\bar \sigma}$ depend on $\xi, T, k^2$ and $\omega^2$. In the spinless case, they reduce to standard densities depending only on $T$ and $\mu$. The ``transverse'' components (those with the subscript $t$) do not appear in the spinless case as they are multiplied by $t^\mu$ that vanishes in the limit $k, \omega \to 0$. 

The energy-momentum decomposition reads
\bel{eq:TmunuG}
T^{\mu\nu} &=& {\bar \varepsilon} u^\mu u^\nu - {\bar P} \Delta^{\mu\nu} + P_k \, k^\mu k^\nu \nn \\ && + P_\omega \,\omega^\mu \omega^\nu + P_t \,(t^\mu u^\nu + t^\nu u^\mu),
\eel
where ${\bar \varepsilon}$ and ${\bar P}$ depend also on $\xi, T, k^2$ and $\omega^2$. We note that \EQ{eq:TmunuG} shares some common features with the energy-momentum tensor used in anisotropic MHD, for example, see Eq.~(39) in~\CITn{PhysRevE.51.4901}. However, one can spot also differences, as the two systems are obviously different. Finally, the form of the spin tensor is given by \EQn{eq:spint_GLW}, which gives 
\bel{eq:Somega}
\f{1}{2} \omega_{\alpha\beta}  S^{\mu, \alpha \beta} &=& u^\mu (A k^2 - A_1 \omega^2) + A_3 t^\mu \nn \\
&\equiv& {\bar s} u^\mu + s_t t^\mu.
\eel 
The use of the above decompositions in \EQn{eq:Hmu2} leads (after comparing the terms multiplying $u^\mu$ and $t^\mu$) to two equations
\bel{eq:scalarG1}
{\bar \varepsilon} + \coth\xi \,{\bar n} T
&=& T {\bar \sigma} + \mu {\bar n} + {\bar s} T, \\
P_t + \coth\xi \, n_t T
&=& T \sigma_t + \mu n_t + s_t T.
\label{eq:scalarG2}
\eel 
In the spinless case, all terms in \EQ{eq:scalarG2} vanish, while \EQ{eq:scalarG1} reduces to \EQ{eq:ext_S} -- the term $\coth\xi \,{\bar n} T$ becomes equal to the equilibrium pressure of spinless particles. 

In this way we arrive at our second main point. Our analysis shows that \EQ{eq:ext_Ss} is not an appropriate starting point for introducing thermodynamics of spin polarized media. We need at least two scalar equations, \EQSTWO{eq:scalarG1}{eq:scalarG2}, to introduce mutual relations between functions describing densities of various physical quantities.

\smallskip
\noindent {\it Non-equilibrium entropy current ---}  To extend the formalism presented above to a theory covering dissipative phenomena, we follow the method initiated by Israel and Stewart~\CITn{Israel:1979wp}. It relies on the replacements of the equilibrium currents $N^\mu$, $T^{\mu \alpha}$ and $S^{\mu, \alpha \beta}$ in \EQ{eq:Hmu2} by the general non-equilibrium expressions that can be represented as the equilibrium terms plus non-equilibrium corrections: $N^\mu_{\rm nq} = N^\mu + \delta N^\mu$, $T^{\mu \alpha}_{\rm nq} = T^{\mu \alpha} + \delta T^{\mu \alpha}$ and $S^{\mu, \alpha \beta}_{\rm nq} = S^{\mu, \alpha \beta} +\delta S^{\mu, \alpha \beta}$, in this way we obtain
\bel{eq:HmuN}
S^\mu_{\rm nq} =  T^{\mu \alpha}_{\rm nq} \beta_\alpha-\f{1}{2} \omega_{\alpha\beta} S^{\mu, \alpha \beta}_{\rm nq}
-\xi N^\mu_{\rm nq} + {\cal N}^\mu.
\eel
Next, we calculate the divergence of the entropy current defined by \EQ{eq:HmuN}. Since in the non-equilibrium case the energy-momentum tensor contains non-symmetric parts, we should use the equations~\footnote{Since only the total angular momentum is conserved in the general case, $\p_\mu J^{\mu, \alpha\beta}_{\rm nq}=0$ with $J^{\mu, \alpha\beta}_{\rm nq} = x^\alpha T^{\mu \beta}_{\rm nq} - x^\beta T^{\mu \alpha}_{\rm nq} + S^{\mu, \alpha\beta}_{\rm nq}$, we have $\p_\mu S^{\mu, \alpha\beta}_{\rm nq}= T^{\beta \alpha}_{\rm nq}-T^{\alpha \beta}_{\rm nq}$. }
\bel{eq:con_eqN}
\!\!\!\!\! \p_\mu N^\mu_{\rm nq}=0, \,\,\, 
\p_\mu T^{\mu \nu}_{\rm nq}=0, \,\,\,
\p_\mu S^{\mu, \alpha \beta}_{\rm nq} 
= T^{\beta \alpha}_{\rm nq} - T^{\alpha \beta}_{\rm nq} .
\eel
This leads to the following expression for the entropy production
\bel{eq:divS}
\p_\mu S^\mu_{\rm nq} &=&  
- \delta N^\mu_{\rm nq} \p_\mu \xi
+ \delta T^{( \mu \lambda )}_{\rm nq} \p_\mu \beta_\lambda \\
&&
+ \delta T^{[ \mu \lambda ]}_{\rm nq} \LB \p_\mu \beta_\lambda 
- \omega_{\lambda\mu} \RB
-\f{1}{2} \delta S^{\mu, \alpha \beta}_{\rm nq} \p_\mu \omega_{\alpha\beta} , \nn
\eel
where the round (squared) brackets denote the symmetric (antisymmetric) parts of the energy-momentum tensor.  

Equation \EQn{eq:divS} implies that the {\it global equilibrium} is defined by the generalized Tolman-Klein conditions~\CITn{Tolman:1934,Klein:1949} which include the two standard equations, $\p_\mu \xi = 0$ and $\p_{( \mu} \beta_{\lambda )} = 0$, and an extra constraint that the spin polarization tensor is given by the thermal vorticity, $\omega_{\lambda\mu} = \p_{[ \mu} \beta_{\lambda ]}$. Nevertheless, in any state that is different from global equilibrium, including the local equilibrium considered herein, the tensors $\omega_{\lambda\mu}$ and $\p_{[ \mu} \beta_{\lambda ]}$ are not directly related and may be significantly different from each other. This behavior is similar to the behavior of the ratio $\xi = \mu/T$. In global equilibrium $\xi =$~const, while in local equilibrium a direct connection between $T$ and $\mu$ is lost. 

Although \EQ{eq:divS} or its special case with $\xi=0$ was obtained before (see, for example: Eq.~(10) in \CITn{Hattori:2019lfp}, (23) in \CITn{Fukushima:2020ucl}, (21) in \CITn{Biswas:2023qsw}, and the QFT discussion in \CITn{Becattini:2023ouz}), the previous studies considered always the local equilibrium state without the orthogonal corrections discussed above. Thus, it is important to extend the previous analyses by considering a different reference point for local equilibrium quantities. For the baryon current one finds
\bel{eq:Ndec}
N^\mu_{\rm nq} = {\bar n}(T,\xi, k^2, \omega^2) u^\mu  + n_t(T,\xi) t^\mu - \lambda \nabla^\mu \xi ,
\eel
where $\lambda \geq 0$ is the diffusion coefficient. Similar expressions although more complicated and lengthy can be found for the energy-momentum and spin tensors. They will be presented and discussed in a separate paper.

\noindent {\it Summary and conclusions ---} In this work we have introduced generalized thermodynamic relations into the framework of a relativistic perfect spin hydrodynamics. They allow for a consistent treatment of spin degrees of freedom, including the use of spin tensors whose structure follows from microscopic calculations.  Our main observation that a commonly used scalar version of thermodynamic relations should be replaced by the tensor form is very general --- the spin hydrodynamics introduces a new hydrodynamic variable that has a tensor structure, hence, in local equilibrium all currents and tensors may a priori have a richer structure than that used in spinless hydrodynamics. To large extent, this situation is similar to the case of MHD. In the Supplemental Material we demonstrate that the same form of the generalized thermodynamic relations (Eqs.~\EQn{eq:Hmu2}, \EQn{eq:firstlaw_TsG}, and \EQn{eq:GD_TsG}) is obtained for the FD statistics, which again supports a universal character of our results.

The obtained results are crucial for establishing consistency between different formulations of spin hydrodynamics. They also form a suitable starting point for introducing dissipative corrections. In the future investigations, it will be useful to establish relations between our results and other works that aim at the construction of a non-equilibrium entropy current. In a recent paper that uses quantum statistical methods~\CITn{Becattini:2023ouz}, a similar structure of the entropy current to ours is obtained. Moreover, the entropy production formula found in~\CITn{Becattini:2023ouz} agrees with the IS ansatz used in this work. On the other hand, general considerations presented in~\CITn{Becattini:2023ouz} do not touch upon the importance of using tensor forms of thermodynamic relations, which is the main issue discussed in this work. We find~\CIT{Becattini:2023ouz}  and our work as complementary analyses -- the first can be treated as the top-down approach, while the latter as the bottom-up method.  Eventually, the two frameworks should converge in practical applications of spin hydrodynamics to describe the experimental data.

\noindent
{\it Acknowledgements ---} The authors thank Samapan Bhadury, Zbigniew Drogosz, and Radoslaw Ryblewski for clarifying discussions. This work was supported in part by the Polish National Science Centre Grant No. 2022/47/B/ST2/01372.

\bibliography{spin-lit}

\newpage

\begin{widetext}
\begin{center}
{\bf SUPLEMENTAL MATERIAL}
\end{center}

Herein, we discuss in more detail the state of local thermodynamic equilibrium as defined within the framework of kinetic theory describing particles with spin~$\onehalf$. A combination of the two concepts: local equilibrium for particles with spin and the conservation laws for the baryon number, energy, linear momentum and spin part of the angular momentum leads to the framework of perfect spin hydrodynamics. 

\section{1. Classical spin description}

    In the classical treatment of spin~\CITn{Mathisson:1937zz,2010GReGr..42.1011M}, one introduces the particle internal angular momentum tensor $s^{\alpha\beta}$ defined by the formula
\bel{eq:sab}
s^{\alpha\beta} = \f{1}{m} \epsUabgd p_\gamma s_\delta.
\eel
Here $p$ is the particle four-momentum satisfying the on-mass-shell condition $p^\mu p_\mu = m^2$ (with $m$ being the particle mass) and $s$ is the particle spin four-vector. Equation \EQn{eq:sab} implies that $s^{\alpha\beta} = -s^{\beta\alpha}$ and $s^{\alpha\beta} p_\beta = 0$. The spin four-vector is orthogonal to four-momentum $s \cdot p = 0$, hence we can write
\bel{eq:sa}
s^{\alpha} = \f{1}{2m} \,\epsUabgd p_\beta s_{\gamma \delta}.
\eel%
In the particle's rest frame (PRF), where $p^\mu = (m,0,0,0)$, the four-vector $s^\alpha$ has only space components, $s^\alpha = (0,\sv_*)$, with the normalization $\spinl = \spin$. For particles with spin $\onehalf$, we use the value of the Casimir operator to set $\spin^2 = \onehalf \left( 1+ \onehalf  \right) = \threefourths$.

The basic object used in the kinetic theory is the phase-space distribution function $f(x,\pv)$. For particles with spin, $f(x,\pv)$ is generalized to a spin dependent distribution $f(x,\pv,s)$. One commonly uses the notation $f(x,p,s)$ for $f(x,\pv,s)$ remembering that the energy $p^0$ is on the mass shell, namely, $p^0 = E_p =\sqrt{m^2 + \pv^2}$. Alongside with the distribution function we introduce the integration measures in momentum  and spin spaces~\CITn{Florkowski:2018fap}
\bel{eq:dP_dS}
dP = \f{d^3p}{(2\pi)^3 E_p}, \hspace{0.4cm} dS = \f{m}{\pi \spin}  \, d^4 s \, \delta(s \cdot s + \spin^2) \, \delta(p \cdot s).
\eel

The two delta functions control here the normalization of the spin vector and its orthogonality to particle momentum. The prefactor in \EQn{eq:dS} is chosen to yield the normalization condition
\bel{eq:dS2}
\int dS = 2
\eel
that reflects two possible orientations of the spin $\onehalf$. Further useful integrals are~\CITn{Florkowski:2018fap}:
\bel{eq:dSs} 
\int dS \,s_{\alpha} &=& 0, \hspace{0.4cm}
\int dS \,s_\sigma s_\rho = 
-\f{2 \spin^2 }{3}\LB g_{\sigma \rho} + p_\sigma p_\rho \RB. \label{eq:dSss}
\eel
They can be used to derive the other two relations that can be frequently used below
\bel{eq:dSos} 
\int dS \, \omega : s &=& 0,  \\
\int dS \, s^{\mu\nu} \, \omega : s &=&    \f{4 \spin^2 }{3 m^2}
\LSB m^2  \omega^{\mu \nu} + p_\alpha \LB p^\mu \omega^{\nu\alpha}
-p^\nu \omega^{\mu\alpha} \RB \RSB
,  \label{eq:dSsos} \\
\int dS \, (\omega : s)^2  &=& 
\f{4 \spin^2 }{3 m^2}
\LB m^2  \omega : \omega + 2 \, p^\alpha p^\beta \omega^{\gamma}_{\,\,\, \alpha} \omega_{\beta\gamma} \RB. \label{eq:dSosos}
\eel
%

\section{2. Fermi-Dirac equilibrium distribution functions -- macroscopic currents}

 In local equilibrium, the spin dependent distribution functions for particles ($+$) and antiparticles ($-$) have the Fermi-Dirac form
\bel{eq:fpm-FD}
f^{\pm_{\rm }}(x,p,s) = \LSB
\exp \LB
\mp \xi(x) + p \cdot \beta(x)  -  \frac{1}{2} \, \omega(x) : s \RB +1 \RSB^{-1}.
\eel 
We also use the compact notation
\bel{eq:fpm-FDy}
f^{\pm_{\rm }} = \f{1}{e^{y^\pm} + 1}
\eel 
with
\bel{eq:ypm}
y^\pm = \mp \xi(x) + p \cdot \beta(x) 
- \frac{1}{2} \omega(x) : s.
\eel

The macroscopic currents and tensors are obtained as moments of the distribution functions. In this way we obtain the baryon current
\bel{eq:Nmu}
N^\mu\!=\!\int dP \,dS \, p^\mu \, \left[f^+(x,p,s)-f^-(x,p,s) \right],
\eel
the energy-momentum tensor
\bel{eq:Tmunu}
T^{\mu \nu}\!=\!\int dP \,dS \, p^\mu p^\nu \, \left[f^+(x,p,s) + f^-(x,p,s) \right],
\eel 
and the spin tensor
\bel{eq:Slmunu}
\hspace{-0.5cm}S^{\lambda, \mu\nu}\!&=&\!\!\int \!dP \, dS \, \, p^\lambda \, s^{\mu \nu} 
\left[f^+(x,p,s)+ f^-(x,p,s) \right].
\eel
In addition we define the current
\bel{eq:CNmu}
\mathcal{N}^\mu\!=\!-\int dP \,dS \, p^\mu \, \left[\ln(1-f^+)+\ln(1-f^-) \right].
\eel
In the traditional hydrodynamics $\mathcal{N}^\mu$ can be directly expressed by local pressure and hydrodynamic flow, $\mathcal{N}^\mu = P \beta^\mu$. However, this is not the case for the spin hydrodynamics. Finally, we introduce the entropy current
\bel{eq:Smu_FD}
S^{\mu} &=& -\int dP \,dS \, p^\mu \, \left[f^+ \ln{f^+}-f^+\ln(1-f^+)+\ln(1-f^+)\right] \\&-& \int dP \,dS \, p^\mu \left[ \, f^- \ln{f^-}-f^-\ln(1-f^-)+\ln(1-f^-) \right]. \nonumber
\eel
Using the identity
\bel{eq:nf1}
f \ln f - f \ln(1-f)= -\f{y}{e^{y}+1} = -y f
\eel
one can show that the entropy current $S^\mu$ can be expressed as a linear combination of other tensors and currents. Inserting \EQn{eq:nf1} into \EQn{eq:Smu_FD} we obtain 
\bel{eq:Smu_FD_Comp}
S^\mu 
&=& \int dP \,dS \, p^\mu \LSB f^+ y^+ - f^- y^- \RSB 
-  \int dP \,dS \, p^\mu  \left[\ln(1-f^+)+\ln(1-f^-) \right] \\
&=&  \int dP \,dS \, p^\mu \LSB f^+ \LB  
-\xi +p \cdot \beta - {\scriptstyle{ \frac{1}{2} }}\omega: s \RB + f^- \LB  
\xi +p \cdot \beta - {\scriptstyle{ \frac{1}{2} }}\omega : s \RB   \RSB + \mathcal{N}^\mu \nn
\eel
or
\bel{eq:Smu_FD}
S^\mu = -N^\mu \xi + T^{\mu \alpha}\beta_\alpha - {\scriptstyle{ \frac{1}{2} }}S^{\mu,\alpha \beta}\omega_{\alpha \beta} + \mathcal{N}^\mu.
\eel

\section{3. Generalized thermodynamic relations for the FD case}

For the distribution function of the form \EQn{eq:fpm-FDy}, one finds useful relations:
\bel{}
df \!=\! -\f{ e^y }{ (e^y+1)^2 }dy \!=\! -\f{1}{e^y+1} \cdot \f{e^y+1-1}{e^y+1}dy \\ \!=\! \f{-1}{e^y+1}\LB \f{e^y+1}{e^y+1} -\f{1}{e^y+1} \RB \!=\! -f(1-f)dy, \nn 
\eel 
\bel{}
d \ln{f} \!=\! \f{df}{f} \!=\! -(1-f)dy,
\hspace{0.4cm}
d \ln{(1-f)} \!=\! f\f{1-f}{1-f}dy \!=\! fdy.
\eel
They can be used to obtain the expression for $d\mathcal{N}^\mu$ directly from \EQn{eq:CNmu}, namely,
\bel{eq:CNmu_FD_CL}
&& d\mathcal{N}^{\mu} 
= -d\int dP \,dS \, p^\mu \, \left[\ln(1-f^+)+\ln(1-f^-) \right]  \\
&& = - \int dP \,dS \, p^\mu \, \LSB f^{+}(-d\xi +p \cdot d\beta - {\scriptstyle{ \frac{1}{2} }} d\omega : s) + f^{-}(d\xi +p \cdot d\beta - {\scriptstyle{ \frac{1}{2} }} d\omega : s)\RSB \nonumber \\ 
&& =
\int dP \,dS \, p^\mu \, (f^+\!-\!f^-)d\xi-\!\!\!\int dP \,dS \, p^\mu p^\alpha (f^+\!+\!f^-) d\beta_{\alpha} + \frac{1}{2} \int dP \,dS \, p^\mu s^{\alpha\beta}(f^+\!+\!f^-)d\omega_{\alpha \beta}, \nonumber 
\eel 
which gives
\bel{eq:dcalN}
d\mathcal{N}^{\mu} = N^{\mu}d\xi - T^{\mu \alpha}d\beta_{\alpha} + {\scriptstyle{ \frac{1}{2} }}S^{\mu,\alpha \beta}d\omega_{\alpha \beta}.
\eel
Finally, we use the form \EQn{eq:Smu_FD} to calculate $dS^\mu$. Combining the obtained result with $d\mathcal{N}^{\mu}$ given by \EQn{eq:dcalN} we find
\bel{eq:dS}
dS^{\mu}= -\xi dN^\mu + \beta_{\alpha}dT^{\mu \alpha}- {\scriptstyle{ \frac{1}{2} }}\omega_{\alpha \beta} dS^{\mu,\alpha \beta}.
\eel
%
\section{4. The limiting case of the Boltzmann statistics}

The classical statistics is obtained by neglecting the term $+1$ in~\EQn{eq:fpm-FD}. In this way we obtain the Boltzmann distribution
\bel{eq:fpm-spin-B}
\widetilde{f}^{\pm_{\rm }}(x,p,s) = \exp \LSB \pm\xi(x)-p \cdot \beta(x)
+ \frac{1}{2} \omega(x) : s \RSB. 
  \eel 
In addition, as the classical limit corresponds to a dilute system, we can always neglect $f$ in expressions such as $(1-f)$  in the corresponding tensors. The next step is to expand the spin-dependent part of the exponent in the \EQn{eq:fpm-spin-B}. Thus we obtain
\bel{eq:expansion3B}
\widetilde{f}^\pm = 
\widetilde{f}^\pm_0 \LSB 1  + \frac{1}{2} \, \omega : s  
+ \frac{1}{8} \,  (\omega : s)^2 + \cdots \RSB,
\eel
Where the $\widetilde{f}^{\pm}_{0}$ is defined as
\bel{eq:fpm-spin-less}
 \widetilde{f}^{\pm}_{0}=\exp[\pm\xi(x)-p \cdot \beta(x)].
\eel

\subsection{4.1 Baryon and particle currents}

Factorization of the Boltzmann distribution function into the ``momentum`` and ``spin`` parts makes the calculation of macroscopic quantities quite straightforward. We first do the integrals over the spin degrees of freedom and subsequently integrate over momentum, which leads to the equations expressed in terms of the modified Bessel functions. In the case of the baryon current, including terms up to the second order, we find
\smallskip
\bel{eq:Computation_Nmu}
N^{\mu} 
&=& 2\sinh{\xi}\int dP\, p^{\mu}e^{-p\cdot \beta}  \LSB  2 + \f{1}{8} \int dS (\omega : s )^2 \RSB  
\\
&=& 4\sinh{\xi}\int dP\, p^{\mu}e^{-p\cdot \beta} \LSB \LB 1 +\frac{\spin^{2}}{12} \omega:\omega \RB +\frac{\spin^{2}}{6m^{2}}p^{\alpha}p^{\beta}\omega^{\gamma}_{\phantom{\gamma} \alpha}\omega_{\beta \gamma} \RSB \nn
\\  
&=& 4\sinh{\xi} \LB 1+\frac{ \spin^{2}}{12} \, \omega:\omega \RB \underset{Z^{\mu}}{\underbrace{\int dP\, p^{\mu}e^{-\beta \cdot p}}}  +
 \frac{2 \textgoth{s:}^{2}\sinh{\xi}}{3 m^{2}}\underset{Z^{\mu \alpha \beta}}{\underbrace{\int dP\, p^{\mu} p^{\alpha} p^{\beta}e^{-\beta \cdot p}}} \omega^\gamma_{\phantom{\gamma}\alpha} \omega_{\beta \gamma}  . \nn
\eel
The integration over the spin degrees of freedom is done according to the rules \EQSTWOn{eq:dSos}{eq:dSosos}, see~\CIT{Florkowski:2018fap}. In the last line of~\EQ{eq:Computation_Nmu}, we have underlined the integrals that define the tensors $Z^\mu$ and $Z^{\mu \alpha \beta}$, whose explicit forms can be found in \CIT{Cercignani:2002rh}~\footnote{Note that the expressions given in \CIT{Cercignani:2002rh} should be divided by $(2\pi)^3$, since our integration measure $dP$ includes this extra factor in the denominator, compare~\EQ{eq:dP_dS}.}
\bel{eq:Zmu}
Z^\mu = \f{T^3}{2\pi^2} z^2  K_2(z) u^\mu,
\eel
\bel{eq:Zmuab}
Z^{\mu\alpha\beta} = -\f{T^5}{2\pi^2} z^3  \LSB 
K_3(z) \LB g^{\mu\alpha} u^{\beta} + g^{\mu\beta} u^{\alpha} + g^{\beta\alpha} u^{\mu} \RB - z K_4(z) u^\mu u^\alpha u^\beta 
\RSB,
\eel
Here $K_n(z)'s$ are the modified Bessel functions of the second type with the argument $z = m/T$. The intermediate steps of the calculations include the contractions:
\bel{eq:RN1}
R^\mu_{N1} = \LB g^{\mu\alpha} u^{\beta} + g^{\mu\beta} u^{\alpha} + g^{\beta\alpha} u^{\mu} \RB \omega^\gamma_{\phantom{\gamma}\alpha} \omega_{\beta \gamma} = \LB 2 \omega^2 - 4 k^2 \RB u^\mu + 2 t^\mu,
\eel
\bel{eq:RN2}
R^\mu_{N2} = u^\mu u^\alpha u^\beta \omega^\gamma_{\phantom{\gamma}\alpha} \omega_{\beta \gamma} =- k^2  u^\mu,
\eel
and
\bel{eq:Z3oo}
Z^{\mu \alpha \beta} \omega^\gamma_{\phantom{\gamma}\alpha} \omega_{\beta \gamma} = -\f{T^5}{2\pi^2} z^3 \LSB K_3(z) R^\mu_{N1} - z K_4(z) R^\mu_{N2} \RSB.
\eel 
This leads to the decomposition
\bel{eq:NmuG}
N^\mu = {\bar n} u^\mu + n_t t^\mu = ( n_{0}+ n_{2}^{k} +  n_{2}^{\omega} )u^{\mu}+n_{t}t^{\mu},
\eel
where the coefficients $n_{0},  n_{2}^{k},  n_{2}^{\omega}$, and $n_t$ have forms:
\bel{eq:n}
n_{0}=\f{2\sinh{\xi}}{\pi^{2}}z^{2}T^{3}K_{2}(z), \hspace{0.4cm}  
{n}_{2}^{k} =-\f{2 \spin^{2} \sinh{\xi}}{3\pi^{2}}   z T^{3}K_{3}(z)k^{2}, 
\eel
\bel{eq:nt}
{n}_{2}^{\omega} = -\f{\spin^{2}\sinh{\xi}}{3\pi^{2}} z T^{3} \LSB z K_{2}(z) + 2 K_{3}(z) \RSB \omega^{2}, \hspace{0.4cm}   n_{t}=-\frac{2\spin^{2}\sinh{\xi}}{3\pi^2}  z T^{3} K_{3}(z).
\eel
The coefficients $n_0$ describes the baryon density of a relativistic spinless gas.

In the case of classical statistics, the current $\mathcal{N}^\mu$ becomes the sum of particle and antiparticle currents, namely
\bel{eq:BNmu}
\mathcal{N}^\mu\!=\!\int dP \,dS \, p^\mu \, \left[\widetilde{f}^+(x,p,s)+\widetilde{f}^-(x,p,s) \right],
\eel
Hence, a simple relation holds
\bel{eq:calN}
{\cal N}^\mu = \coth\xi\,\,N^\mu .
\eel
We note that throughout this work we assume that $\mu \neq 0$ ($\xi \neq 0)$.

\subsection{4.2 Energy momentum tensor}

Using the definition of the energy-momentum tensor \EQn{eq:Tmunu} and expanding the spin part of the distribution functions up to the second order in the spin polarization tensor, we obtain the formula
\bel{eq:TmunuZ}
T^{\mu \nu} &=& 4\cosh{\xi} \left( 1 +\frac{\spin^{2}}{12} \omega:\omega \right)\underset{Z^{\mu \nu}}{\underbrace{\int dP\, p^{\mu} p^{\nu}e^{-\beta \cdot p}}}  \nn \\
&& + \frac{2 \spin^{2} \cosh{\xi} }{3 m^{2}} \underset{Z^{\mu \nu \alpha \beta}}{\underbrace{\int dP\, p^{\mu} p^{\nu} p^{\alpha} p^{\beta}e^{-\beta \cdot p}}} 
\, \omega^{\gamma}_{\phantom{\gamma}\alpha} \omega_{\beta \gamma} .
\eel
Here we have underlined the tensors $Z^{\mu\nu}$ and $Z^{\mu\nu\alpha\beta}$, whose explicit forms are~\CITn{Cercignani:2002rh}:
\bel{eq:Zmunu}
Z^{\mu\nu} = -\f{T^4}{2\pi^2} z^2  \LSB K_2(z) g^{\mu\nu} - z K_3(z) u^\mu u^\nu \RSB
\eel
and
\bel{eq:Zmunuab}
Z^{\mu\nu\alpha\beta} &=& \f{T^6}{2\pi^2} z^3  \LSB 
K_3(z) \LB g^{\mu\nu} g^{\alpha\beta} + g^{\mu\alpha} g^{\nu\beta} + g^{\alpha\nu} g^{\mu\beta} \RB \right. \\ 
&& - z K_4(z) \LB g^{\mu\nu} u^{\alpha} u^{\beta} +  g^{\mu\alpha} u^{\nu} u^{\beta} + g^{\alpha\nu} u^{\mu} u^{\beta} + g^{\mu\beta} u^{\alpha} u^{\nu} + g^{\beta\alpha} u^{\nu} u^{\mu} + g^{\beta\nu} u^{\mu} u^{\alpha} \RB \nn \\
&& \left. + z^2 K_5(z) u^{\mu} u^{\nu} u^{\alpha} u^{\beta} 
\RSB. \nn
\eel
The useful tensor contractions in this case are:
\bel{eq:RT1}
R_{T1}^{\mu\nu} &=& \LB g^{\mu\nu} g^{\alpha\beta} + g^{\mu\alpha} g^{\nu\beta} + g^{\alpha\nu} g^{\mu\beta} \RB \omega^{\gamma}_{\phantom{\gamma}\alpha} 
\omega_{\beta \gamma} \nn \\
&=& 2 \LSB
g^{\mu\nu} (2\omega^2-k^2)-u^\mu u^\nu (k^2+\omega^2) - (k^\mu k^\nu +\omega^\mu \omega^\nu) + u^\mu t^\nu + u^\nu t^\mu 
\RSB,
\eel
\bel{eq:RT2}
R_{T2}^{\mu\nu} &=& \LB g^{\mu\nu} u^{\alpha} u^{\beta} +  g^{\mu\alpha} u^{\nu} u^{\beta} + g^{\alpha\nu} u^{\mu} u^{\beta} + g^{\mu\beta} u^{\alpha} u^{\nu} + g^{\beta\alpha} u^{\nu} u^{\mu} + g^{\beta\nu} u^{\mu} u^{\alpha} \RB \omega^{\gamma}_{\phantom{\gamma}\alpha} 
\omega_{\beta \gamma} \nn \\
&=& -k^2 g^{\mu\nu} + 2 (\omega^2 - 3 k^2) u^\mu u^\nu + 2 (u^\mu t^\nu + u^\nu t^\mu),
\eel
\bel{eq:RT3}
R_{T3}^{\mu\nu} &=& u^{\mu} u^{\nu} u^{\alpha} u^{\beta}  \omega^{\gamma}_{\phantom{\gamma}\alpha} 
\omega_{\beta \gamma} = -k^2 u^\mu u^\nu,
\eel
and
\bel{eq:Z4oo}
Z^{\mu\nu\alpha\beta} \omega^\gamma_{\phantom{\gamma}\alpha} \omega_{\beta \gamma} &=& \f{T^6}{2\pi^2} z^3  \LSB 
K_3(z) R_{T1}^{\mu\nu} - z K_4(z) R_{T2}^{\mu\nu} + z^2 K_5(z) R_{T3}^{\mu\nu}
\RSB. \nn
\eel
Combining all the expressions derived above, we find the formula
\bel{eq:Tmunu_decomposed}
T^{\mu \nu} &=&
{\bar \varepsilon} u^\mu u^\nu - {\bar P} \Delta^{\mu\nu} + P_k \, k^\mu k^\nu   + P_\omega \,\omega^\mu \omega^\nu + P_t \,(t^\mu u^\nu + t^\nu u^\mu) \\
&=& (\varepsilon_{0}+{\varepsilon}_{2}^{k}+{\varepsilon}_{2}^{\omega})u^{\mu}u^{\nu} - ({P}_{0}+{P}^{k}_{2}+{P}^{\omega}_{2})\Delta^{\mu\nu}+P_{k}k^{\mu}k^{\nu}+P_{\omega}\omega^{\mu}\omega^{\nu}+P_{t}(t^{\mu}u^{\nu}+t^{\nu}u^{\mu}), \nn
\eel
where the coefficient functions read:
\bel{coef:epsilon_0}
{\varepsilon}_{0}=\frac{2\cosh{\xi}}{\pi^{2}}z^{2}T^{4} \LSB z K_{3}(z)-K_{2}(z) \RSB,
\eel
\bel{eq:epsilon_k_w}
{\varepsilon}_{2}^{k} &=& -\frac{2\spin^{2}\cosh{(\xi)}}{3 \pi^{2}}z T^{4}\LSB z K_{2}(z) + 5K_{3}(z) \RSB k^{2}, \\
{\varepsilon}_{2}^{\omega} &=& -\f{\spin^{2}\cosh{(\xi)}}{3\pi^{2}} zT^{4}\LSB z K_{2}(z)+(z^{2}+10)K_{3}(z) \RSB\omega^{2},
\eel

\bel{coef:pressure_0}
{P_{0}}=\frac{2\cosh{\xi}}{\pi^{2}}z^{2}T^{4} K_{2}(z),
\eel

\bel{eq:pressure_k_2}
{P}^{k}_{2}=-\frac{4 \spin^{2}\cosh{\xi}}{3 \pi^{2}}zT^{4}K_{3}(z)k^{2}, \hspace{0.4cm} {P}^{\omega}_{2} = -\frac{\spin^{2}\cosh{\xi}}{3\pi^{2}} zT^{4}\LSB z K_{2}(z)+4K_{3}(z) \RSB\omega^{2},  
\eel

\bel{coef:[pressure_t_2]]}
P_{t}=\f{2 \spin^{2} \cosh{\xi}}{3\pi^{2}}zT^{4} \LSB K_{3}(z)-zK_{4}(z) \RSB, \hspace{0.4cm}  P_{k}=P_{\omega}=-\f{2 \spin^{2}\cosh{\xi}}{3\pi^{2}} z \, T^{4} K_{3}(z).
\eel
Obviously, the quantities $\varepsilon_0$ and $P_0$ correspond to the energy density and pressure of spinless particles, respectively. We also have $P_0 = \coth\xi \, n_0 T$, which is the relativistic version of the Clapeyron equation.

\subsection{4.3 Spin tensor}

In the next step we consider the spin tensor $S^{\lambda,\mu \nu}$ given by \EQ{eq:Slmunu}. Using \EQ{eq:dSsos} we can again express it up the second order in $\omega$ using  the tensors $Z$,
\bel{eq:Slmunu}
S^{\lambda,\mu \nu} &=& 2\cosh{\xi}\int dP\, p^{\lambda}e^{-p \cdot \beta}
\int dS s^{\mu \nu}\LSB 1+\f{1}{2} \omega : s \RSB   
\\ &=&
\cosh{\xi}\int dP\, p^{\lambda} e^{-p \cdot \beta}
\, \int 
dS \, s^{\mu \nu} \,  \omega : s \nonumber 
\\ &=&
\f{4\spin^{2} \cosh{\xi} }{3m^{2}}\LSB m^{2}\omega^{\mu \nu}\underset{Z^{\lambda}}{\underbrace{\int dP\, p^{\lambda} e^{-\beta \cdot p}}} +
\omega^{\nu}_{\phantom{\nu}\alpha}\underset{Z^{\lambda \alpha \mu}}{\underbrace{\int dP\, p^{\lambda}p^{\alpha}p^{\mu} e^{-\beta \cdot p}}} - \omega^{\mu}_{\phantom{\mu}\alpha}\underset{Z^{\lambda \alpha \nu}}{\underbrace{\int dP\, p^{\lambda}p^{\alpha}p^{\nu} e^{-\beta \cdot p}}} \RSB. \nn
\eel 
To obtain the final form, we need the explicit expression for the contraction
\bel{eq:ZZ}
&& Z^{\lambda \alpha \mu}\omega^{\nu}_{\phantom{\nu}\alpha} - Z^{\lambda \alpha \nu}\omega^{\mu}_{\phantom{\mu}\alpha} \\
&& = -\f{T^5}{2\pi^2} z^3 \LSB K_3(z) \LB g^{\lambda\alpha} u^{\mu} + g^{\lambda\mu} u^{\alpha} + g^{\mu\alpha} u^{\lambda} \RB - z K_4(z) u^\lambda u^\alpha u^\mu \RSB \omega^\nu_{\,\,\,\alpha} \nn \\
&& \,\,\,\, + \f{T^5}{2\pi^2} z^3 \LSB K_3(z) \LB g^{\lambda\alpha} u^{\nu} + g^{\lambda\nu} u^{\alpha} + g^{\nu\alpha} u^{\lambda} \RB - z K_4(z) u^\lambda u^\alpha u^\nu \RSB \omega^\mu_{\,\,\,\alpha} \nn \\
&& = -\f{T^5}{2\pi^2} z^3 \LSB K_3(z) \LB 
t^{\lambda\mu\nu}
+ u^{\lambda} \omega^{\nu\mu} - u^{\lambda} \omega^{\mu\nu} \RB  
- z K_4(z) u^\lambda \LB u^\mu k^\nu - u^\nu k^\mu \RB \RSB. \nn
\eel
where
\bel{eq:tlmunu}
t^{\lambda\mu\nu} &=& \omega^{\nu\lambda} u^{\mu} - \omega^{\mu\lambda} u^{\nu}
+ g^{\lambda\mu} k^{\nu} - g^{\lambda\nu} k^{\mu} \nn \\
&=& u^\lambda \LB u^\mu k^\nu - u^\nu k^\mu \RB + u^\mu t^{\nu\lambda} - u^\nu t^{\mu\lambda} +  g^{\lambda\mu} k^{\nu} - g^{\lambda\nu} k^{\mu}.
\eel
Then, the spin tensor $S^{\lambda,\mu \nu}$ can be written as
\bel{eq:Slmunu_final}
S^{\lambda, \mu \nu}=A_{1}u^{\lambda}\omega^{\mu \nu}
+\f{A_{2}}{2} u^{\lambda} \LB u^{\mu}k^{\nu} - u^{\nu}k^{\mu} \RB
+\f{A_{3}}{2} \, t^{\lambda \mu \nu}
\eel
where
\bel{eq:A1_A2_A3}
A_{1} &=& \f{2\spin^{2}\cosh{\xi}}{3\pi^{2}}zT^{3} \LSB zK_{2}(z)+2K_{3}(z)\RSB, 
\\
A_{2} &=& \f{4\spin^{2}\cosh{\xi}}{3\pi^{2}}z^{2}T^{3}K_{4}(z), 
\\
A_{3} &=& -\f{4\spin^{2}\cosh{\xi}}{3\pi^{2}}z T^{3}K_{3}(z) .
\eel
This result is consistent with the decomposition used in \CIT{Florkowski:2019qdp}. It is also convenient to introduce a coefficient $A$ defined by the expression
\bel{eq:A}
A = A_1 - \f{A_2}{2}-A_3, 
\eel
which allows to rewite the spin tensor in a compact form as
\bel{eq:spint_GLW}
S^{\lambda, \mu \nu} = u^\lambda 
\LSB A \LB k^\mu u^\nu - k^\nu u^\mu \RB + A_1 t^{\mu\nu}
\RSB  + \frac{A_3}{2} 
\LB t^{\lambda \mu} u^\nu - t^{\lambda \nu} u^\mu + 
\Delta^{\lambda \mu} k^\nu - \Delta^{\lambda \nu} k^\mu 
\RB.
\eel

\end{widetext}

\end{document}